\documentclass[aps,12pt]{revtex4}

\usepackage{graphicx}
\usepackage{graphics}
\usepackage{color}
\usepackage[intlimits,tbtags]{amsmath}
\usepackage{amsfonts,amssymb,amsthm}

\def\beq{\begin{equation}}
\def\eeq{\end{equation}}
\def\bea{\begin{eqnarray}}
\def\eea{\end{eqnarray}}

\begin{document}

\title{One-dimensional irreversible aggregation with TASEP dynamics}

\author{N. Zh. Bunzarova~$^{\dag^1\dag^2}$  and N. C. Pesheva~$^{\dag^2}$}

\affiliation{$^{\dag^1}$~Bogoliubov Laboratory of Theoretical Physics, Joint Institute for Nuclear Research, 141980 Dubna, Russia}

\affiliation{$^{\dag^2}$~Institute of Mechanics, Bulgarian
Academy of Sciences, 1113 Sofia, Bulgaria}

\email{nadezhda@imbm.bas.bg}

\begin{abstract}

We define and study one-dimensional model of irreversible aggregation of particles obeying a discrete-time kinetics, which is a special limit of the  generalized Totally Asymmetric Simple Exclusion Process (gTASEP) on open chains. The model allows for clusters of particles to translate as a whole entity one site to the right with the same probability as single particles do. A particle and a cluster, as well as two clusters,
irreversibly aggregate whenever they become nearest neighbors. Nonequilibrium stationary phases appear under the balance of injection and ejection of particles. By extensive Monte Carlo simulations it is established that the phase diagram in the plane of the injection-ejection probabilities consists of three stationary phases: a multi-particle (MP) one, a completely filled (CF) phase and a 'mixed' (MP+CF) one. The transitions between these phases are: an unusual transition between MP and CF with jump discontinuity in both the bulk density and the current, a conventional first-order transition with a jump in the bulk density between MP and MP+CF, and a continuous clustering-type transition from MP to CF, which takes place throughout the MP+CF phase between them. By the data collapse method a finite-size scaling function for the current and bulk density is obtained near the unusual phase transition line. A diverging correlation length, associated with that transition, is identified and interpreted as the size of the largest cluster. The model allows for a future extension to account for possible cluster fragmentation.

\end{abstract}

\maketitle

\section{Introduction}

Irreversible aggregation of clusters of arbitrary size arises in many physical-chemical processes as aerosol physics, polymer growth, and
even in astrophysics \cite{Lee00}. The ability to control aggregation of proteins could be an important tool in the arsenal of the drug development.
However, in biochemistry of life this process may play a destructive role as well. For example, many neurodegenerative diseases, including Alzheimer’s disease,
Parkinson’s disease, prion diseases, to mention some, are characterized by intracellular aggregation and deposition of pathogenic proteins \cite{THF02}.
Moreover, the abnormal irreversible aggregation of ribosomes leads to irreparable damage of protein synthesis and results in neuronal death after focal brain ischemia \cite{ZLH}.

Irreversible aggregation of two clusters, $A_j$ and $A_k$, containing $j$ and $k$ particles, respectively, is usually described by the reaction
\beq
A_j + A_k \rightarrow A_{j+k},
\label{reac}
\eeq
with a rate kernel $K(j,k)$ which, generally, depends on the size of both clusters. The physical process is modelled by the special choice of the
kernel $K(j,k)$. In reaction Eq. (\ref{reac}) the fragmentation of clusters is neglected, i.e., the process is considered as
irreversible aggregation. The theoretical studies of this widely spread in nature phenomenon rapidly grew after the formulation of the corresponding set of ordinary (in time) differential equations in the seminal paper by Smoluchowski \cite{Sm1916}. In the case of continuous cluster-size variable the  kinetics of irreversible aggregation can be described by the integro-differential Smoluchowski equation \cite{Sm1917}.
Most often, the particles and clusters are assumed to undergo a Brownian motion in the real three-dimensional space, or in models with a reduced space dimensionality D = 2 or 1. Then, the collision rate between two clusters is given by $K(j,k)c_jc_k$, where $c_j$ is the concentration of clusters $A_j$. Several cases, corresponding to a special form of the kernel were exactly solved, see the review \cite{Ley03} and references therein. For different classes of kernels, Smoluchowski's coagulation equations were used for the description of the kinetics of gelation, e.g., in
\cite{LT82,HEZ83}. Criteria for the occurrence of gelation were derived and critical exponents in the pre- and post-gelation phase were obtained. In general, the Smoluchowski equation was studied for a large class of symmetric and homogeneous with respect to its arguments kernels $K(j,k)$.

Most of the works in the recent decades on the aggregation process were focused on the existence of scaling laws in the long-time limit of the cluster-size probability distribution, see \cite{Ley03}. It was established that at large times the typical cluster size $s(t)$ increases algebraically with time $t$ as $s(t)\propto t^z$, and the time-dependent probability distribution $P(j,t)$ of the cluster size $j$ obtains the scaling form
\beq
P(j,t) = Ws(t)^{-2}\Phi(j/s(t)),
\label{scale}
\eeq
where $W$ is a constant factor, and the scaling function $\Phi$ satisfies a certain integral equation.

Besides the rate equations approach, developed by Smoluchowsky, there appeared statistical studies, based on combinatoric calculations of the aggregate size distribution. As shown by Flory, very large aggregates can appear suddenly at a certain critical extent of the
3D polymerization reaction \cite{F41}. Stockmayer extended Flory's results to branched-chain polymers and argued that the transition from liquid to
gel is analogous to the condensation of a saturated vapor \cite{Stock43}. At the gel point $t = t_c$ the cluster size distribution was shown to have a power-law asymptotic behavior, $c_j(t_c) \simeq Cj^{-\tau}$, $j\rightarrow \infty$, where $c_j(t)$ is the concentration of $j$-mers at time $t$.
The classical Flory-Stockmayer theory predicted $\tau = 5/2$. More recently, critical kinetics near gelation was studied in \cite{LT82, L12}. It was shown, that even starting from initial conditions with monomers only, an infinite cluster appears at arbitrarily small times - the phenomenon was called an 'instantaneous gelation' \cite{L12}. This phenomenon is known to arise in cases when the reaction rates increase rapidly with the cluster size.  It was shown that the asymptotic behavior of the kernel $K(j,k)$ at large values of $j$ and $k$ is of crucial importance for the size, $k$, and time, $t$, asymptotic behavior of the cluster-size probability distribution. The mechanism of the appearance of a power-law tail in the cluster-size distribution at large cluster sizes was investigated in \cite{L12}.

On the ground of experiments on aqueous dispersions of polystyrene models, the authors of \cite{OLSetal04} have concluded that the known by 2004 aggregation-fragmentation models were unable to reproduce the experimental observations. They argued that real aggregates, depending on their shape, may experience anisotropic diffusion, in contrast to monomers. In addition, effects of weak bonds and cluster breakup should be taken into account. Two extended models, one with multiple particle contacts, and the other with an exponentially relaxing sticking probability, were found to better agree with the experimental data.

On the one hand, the one-dimensional cluster kinetics is free from most of the above mentioned complications - all the clusters have the same shape and their sticking probability should be size-independent. On the other hand, the rate equation approach neglects spatial fluctuations in the particle (cluster) concentration, which are expected to be essential in 1D aggregation processes. As shown by Dongen, the Smoluchowski's coagulation equations
lead to incorrect predictions at large times for space dimensions $d \le d_c$, where the upper critical dimension $d_c$ turned out to be model dependent
\cite{Don89}. For example, if one considers only the number of clusters, i.e., in the case of the reaction $A+A \rightarrow A$, then $d_c = 2$. Kang and Redner have shown that the same upper critical dimensions holds for size-independent rate kernel $K(j,k)=1$ and diffusion constant $D$ \cite{KR84}. On the basis of this result and computer simulations, some authors have incorrectly concluded that $d_c = 2$ holds generally in irreversible aggregation, see, e.g., \cite{ME88}.

An exact solution for a diffusion limited polymerization process in 1D was obtained by Spouge \cite{Sp88}. The initial state was assumed to contain monomers only, the initial distances between consecutive monomers being independently and identically distributed. Then, monomers start to diffuse identically and independently in 1D, and aggregate when they meet in pairwise collision (this process is called "Ppoly"). Diffusion on the integer lattice with a drift $d\not= 0$ was also considered and the same solution was found as in the driftless case, but with a diffusion constant 2$D +d$ instead of 2$D$. The expected concentration of $k$-mers at large times $t$, was shown to decay as $c_k(t)\propto t^{-3/2}$. A diffusion-limited single-species irreversible aggregation process $A + A \rightarrow A$ in 1D, with random particle input in the bulk, was suggested and exactly solved for the steady state in \cite{DbA89}. The results show that no autonomous first-order rate equation can describe the macroscopic behavior of the system.
Another, exactly solved in one dimension aggregation model, where particles are growing by heterogeneous condensation, i.e., when aggregation takes place only on existing particles involved in Brownian motion, without forming new nuclei, and particles merge upon collision, was proposed in \cite{HH08}. The kinetics involved in this model violates the conservation of mass law. The analytical solution of the model was obtained by using a generalized Smoluchowski equation, including the velocity with which particles grow by condensation. As a result of the additional growth by condensation, the sizes of the colliding particles are increased by a fraction $\alpha$ of their respective sizes, and the following law for algebraic growth of the mean cluster size in the long-time limit
 was found to read, $s(t) \propto t^{1+2\alpha}$, instead of the linear growth $s(t) \propto t$ in the absence of condensation.

A statistical thermodynamics of clustered populations ($M$ particles distributed into $N$ clusters) was presented by Mitsoukas \cite{M14}. The emergence of a giant component (gel phase) was treated as a formal phase transition and a thermodynamic criterion for its appearance was formulated in a way analogous to the case of systems in equilibrium.

Our aim here is to propose and study a new discrete-time stochastic model of irreversible aggregation of hard-core particles on open chains.  In particular, the model should admit detailed study of fluctuations and finite-size effects, both in the time evolution of the initial state and in the non-equilibrium stationary states induced by the boundary conditions. To this end, we implement  a  discrete TASEP-like dynamics with the following properties: (i) existing clusters of particles are never fragmented into parts, (ii) clusters are translated as a whole entity one site to the right, provided the target site is empty,  with the  same probability as single particles do, and (iii) any two particles or clusters, occupying consecutive positions on the chain, may become nearest-neighbors and aggregate irreversibly into a single cluster. We show that a model with the above properties can be obtained as a special  limit of the generalized totally asymmetric simple exclusuion process (gTASEP). The gTASEP has been recognized as an exactly solvable model by W\"{o}lki \cite{W05}. However, it was defined and studied under
periodic boundary conditions only, see \cite{DPPP, DPP15,AB16}. To study boundary driven stationary states of open finite systems, we set the left boundary condition as it follows in the corresponding limit of the one for the gTASEP, which we define in Sec.~II.  Since our model has not been solved exactly, our main aim here is to investigate its stationary phases and the nonequilibrium transitions between them, mainly by means of extensive Monte Carlo simulations.

It is in place here to mention that the asymmetric simple exclusion process (ASEP) is one of the simplest exactly solved models of driven many-particle with particle conserving bulk stochastic dynamics, see the reviews \cite{D98,S01}. In the extremely asymmetric case, when particles are allowed to move in one direction only, it reduces to the TASEP. For description of the model in the context of interacting Markov processes we refer to \cite{S70}. Presently, ASEP and TASEP are paradigmatic models for understanding a variety of nonequilibrium phenomena. Devised to model kinetics of protein synthesis \cite{MGP68}, TASEP and its numerous extensions have found many applications to vehicular traffic flow \cite{CSS00,H01,Schad01}, biological transport \cite{CSN05,PK05,NKP13,BPB14,TKM15,CTK15}, one-dimensional surface growth \cite{KS88,S05}, forced motion of colloids in narrow channels \cite{CL99,K07}, spintronics \cite{RFF07}, current through chains of quantum dots \cite{KO10}, etc. Notably, traffic-like collective motion is found also at different levels of all biological systems. In particular, it was noted that when molecular motors, like kinesin and dynein, encounter a traffic jam along their 1D track of transport in the cells, the result is a disease of the living being \cite{CSN05,PK05}. It is the effect of the particular property of the 1D TASEP-like dynamics, which blocks any overtaking and exhibits spontaneous traffic jams in high density flow, that we want to study in the context of the irreversible aggregation phenomena.

The first exact solution of the original continuous-time TASEP was based on a recurrence relation, obtained at special values of the model parameters in \cite{DDM92}, and then was generalized to the whole parameter space by Sch\"{u}tz and Domany \cite{SD93}. An effective way to exploit the recursive properties of the steady states of various one-dimensional processes offered the matrix-product ansatz (MPA). A matrix-product representation of the steady-state probability distribution for TASEP was found by Derrida, Evans, Hakim, and Pasquier \cite{DEHP}. Their formalism involves two square matrices, generically infinite-dimensional, which satisfy a quadratic algebra known as the DEHP algebra. Krebs and Sandow \cite{KS97} proved that the stationary state of any one-dimensional system with random-sequential dynamics involving nearest-neighbor hopping and single-site boundary terms can always be written in a matrix-product form. The MPA marked a breakthrough in the solution of TASEP in discrete time under periodic as well as open boundary conditions. The general case of ASEP with stochastic sublattice-parallel dynamics was solved in \cite{HP97}. Next, the TASEP with ordered-sequential update was solved by mapping the corresponding algebra onto the DEHP algebra \cite{RSS96,RS97}. Finally, the case of parallel update was solved by using two new versions of the matrix-product ansatz, see \cite{ERS99} and \cite{dGN99}. In general, the MPA has become a powerful method for studying stationary states of different one-dimensional Markov processes out of equilibrium \cite{BE07}.

It should be noted that the properties of the TASEP depend strongly on the choice of the boundary conditions, similarly to the case of systems with long-range interactions.  The open system exhibits (in the thermodynamic limit) three stationary phases in the plane of particle input-output rates, with continuous or discontinuous in the bulk density transitions between them. We emphasize that in our study the fragmentation processes, which in real-life phenomena become increasingly important as clusters grow large, are completely ignored. These fragmentation processes, when taken into account, can lead to the establishment of a stationary state in the system, see, e.g., \cite{ME88}. Instead, we consider finite open chains, where stationary phases are attained in the long-time limit due to the balance between injection and ejection of particles.

The paper is organized in seven sections. In Sec.~II we formulate the model, Sec.~III presents the phase diagram in the plane of particle injection and ejection probabilities. The different stationary nonequilibrium phases are distinguished on the basis of numerically evaluated local density profile, particle current and the probability of complete chain filling. In Sec.~IV we study the transitions between the nonequilibrium stationary phases: we observe an unusual transition with jump discontinuity in both the bulk density and the current, a conventional first-order one with jump in the bulk density only, and a continuous clustering-type transition taking place throughout a whole 'mixed' phase. In Sec.~V, from simple stationarity conditions, we derive exact analytic expressions for the local particle density at the chain ends in the multi-particle phase. On the basis of our numerical data obtained in the neighborhood of the unusual phase transition, in Sec.~VI a finite-size scaling function for the current and the bulk particle density is suggested, and its parameters are evaluated. This scaling function is used for the evaluation of the thermodynamic jumps in the current and bulk density at the transition point, as well as for the identification of a related divergent correlation length. Finally, a summary of the results and some perspectives for further investigations are given in Sec.~VII.

\section{The Model}

\begin{figure}[ht]
\includegraphics[width=140mm]{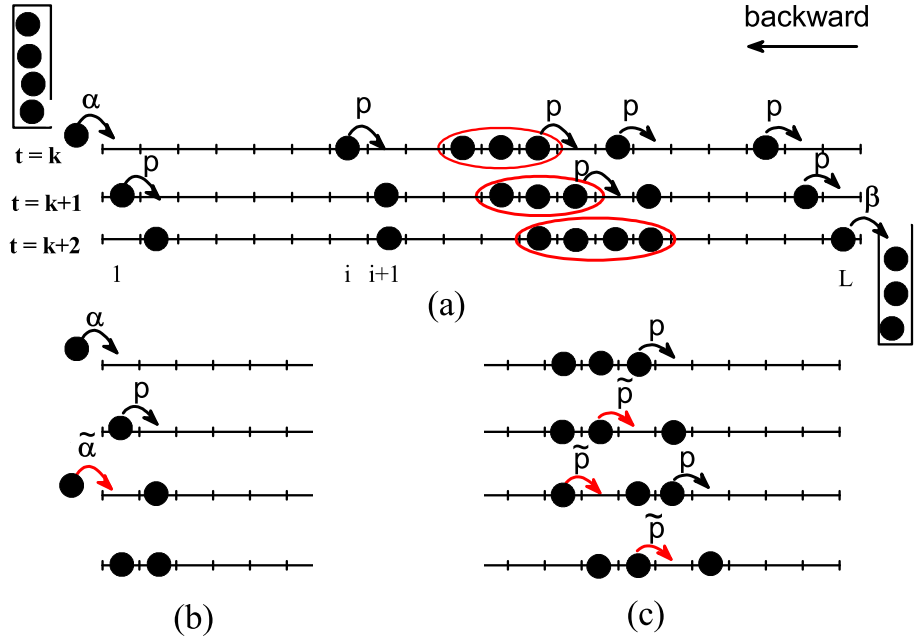} \caption{(Color online) A sketch of the algorithm of our one-dimensional model of irreversible aggregation.
The bulk hopping rules are illustrated in panel (a): a single particle as well as a whole cluster of particles hop one site on the right with
 probability $p$, provided the target site is vacant. Such a hop is shown by black curved arrow with 'p' above it. The aggregation of a cluster with three particles with a single particle is shown to take place at time moments $t = k+1,k+2$. Panel (b) illustrates the left boundary condition (see text). Panel (c) illustrates the particle hopping rules in the generalized TASEP: black arrows with 'p' above show hopping with probability $p$, and red arrows with '$\tilde{\rm p}$' above it - a hop with the modified probability $\tilde{p}$. Our model corresponds to the limit case of  $\tilde{p} = 1$.}   \label{Sketch}
\end{figure}

To define our model, we start with reminding the reader the bulk kinetics of the generalized TASEP, \cite{DPPP, DPP15,AB16}, and adapt it to open chains. Consider an open chain of $L$ sites, labeled from the left to the right by $i = 1,2,\dots , L$. The sites can be empty or occupied by just one particle.  During each moment of discrete time $t_k$, $k=0,1,2,\dots$, the configuration of the whole chain takes place in $L+1$ consecutive steps in a backward sequential order.  First, if the last site $L$ is occupied, the particle is removed from it with probability $\beta$, and stays in place with probability $1-\beta$. Next, all the pairs of nearest-neighbor sites are updated in the order $(L-1,L), \dots, (i,i+1),\dots, (1,2)$. At that, the probability of a hop along a bond $(i,i+1)$ depends on whether a particle has jumped from site $i+1$ to site $i+2$, when the bond $(i+1,i+2)$ was updated at the same moment of time, or not. If the particle at site $i$  is the first (rightmost) particle of a cluster, or it is isolated, then it hops to an empty site $i+1$ with probability $p$, and stands immobile with probability $1-p$. If a particle at $i+1$ belongs to a cluster and has hopped forward to site $i+2$, thus leaving site $i+1$ empty, then the  particle at site $i$ from the same cluster hops to site $i+1$ with a modified probability  $\tilde{p}$ and stays immobile with probability $1-\tilde{p}$, see Fig. \ref{Sketch}.  In the model with the above generalized backward-ordered dynamics, called gTASEP, a cluster of $k$ particles is translated in the bulk as a whole entity by one site to the right with probability $p\tilde{p}^{k-1}$, and is fragmented into two parts with probability $p(1-\tilde{p}^{k-1})$. Finally, the first site of the chain has to be updated. Suppose it is updated in the standard TASEP way: if the first site is empty, a particle is injected in the system with probability $\alpha$, and the site remains empty with probability $1-\alpha$ . Then, the bulk kinetics of the usual TASEP with parallel update is recovered when $\tilde{p} = 0$, but the left boundary condition remains different: under the parallel update, if site $i =1$ was occupied at the beginning of the discrete-time update, then no particle can enter at it. A simple way to restore the rules of the parallel TASEP is to modify the injection probability, by setting it to $\tilde{\alpha} = \min\{\alpha \tilde{p}/p, 1\}$  in the case when the first site was occupied at the beginning of the integer-time moment, but became vacant after the update of the bond $(1,2)$, see panel (b) in Fig. \ref{Sketch}. Obviously, in the case of parallel update $\tilde{p} = 0$  implies  $\tilde{\alpha} =0$. Moreover, the case of the ordinary backward-ordered sequential update is completely recovered when  $\tilde{p} = p$.

Now, it is easily seen that a model with the desired kinetics of irreversible aggregation follows from the above defined gTASEP on open chains in the limit   $\tilde{p} = 1$. Then, the modified injection probability becomes $\tilde{\alpha} = \min\{\alpha/p, 1\}$.

Here we summarize the main features of the kinetics of our model:

(1) When the last site $L$ is occupied, the particle is removed from it with probability $\beta$, and stays in place with probability $1-\beta$.

(2) When the site $i+1$, $i=1,2,\dots , L-1$,  has not changed its occupation number, the probabilities are
the standard ones: if site $i+1$ remains empty, then the jump of a particle from
site $i$ to site $i+1$ takes place with probability $p$, and the particle stays immobile with
probability $1-p$; if site $i+1$ remains occupied, no jump takes place and the configuration of
the bond $(i,i+1)$ is conserved.

(3) If in the previous step of the same configuration update a particle has jumped from site $i+1$ to site $i+2$, thus leaving $i+1$ empty,
then the jump of a particle from site $i$ to site $i+1$ in the next step takes place deterministically (with probability 1).
Thus, no particle can chip off an existing cluster during its translation.

(4) The first site is updated by applying a modified left boundary condition: a particle is injected at the first site of the chain with probability $\alpha >0$,  if the site was vacant at that moment of time, or with probability $\tilde{\alpha} = \min\{\alpha/p, 1\}$, if the site was initially occupied but became vacant after its update at the same moment of time. We emphasize, that a different choice of this boundary condition can change the appearance of the phase diagram but the property of irreversible aggregation, which arises from the bulk dynamics, will persist.

Thus: (i) If the first (rightmost) particle of a cluster moves, all the remaining particles follow it deterministically and, as a result, the position
of the whole cluster is shifted one site to the right; (ii) When $\alpha \ge p$, the stationary state of the system represents a completely aggregated phase, consisting of a single cluster with the size of the chain. In this case the current of particles $J$ equals the ejection probability $\beta$.

Our model is defined as driven lattice gas evolving in discrete time and discrete space which greatly facilitates its theoretical and computational study. The equivalence of the kinetics to a special limit case of the gTASEP gives hope that a future solution for the stationary states of gTASEP on open chains would provide a theoretical explanation of the unusual features of the present model.

\section{Phase diagram}

\begin{figure}[ht]
\includegraphics[width=120mm]{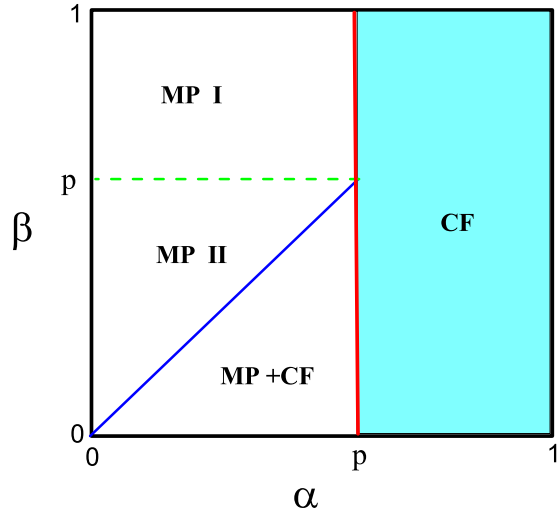} \caption{(Color online) Phase diagram in the plane of injection ($\alpha$) - ejection ($\beta$) probabilities.
The many-particle phase MP occupies two regions, MP~I and MP~II; it contains a macroscopic number of particles or clusters of size $O$(1) as $L\rightarrow \infty$; MP~I and MP~II differ only by the shape of the local density profile (see text). The phase MP+CF is mixed in the sense that its configurations contain with nonvanishing probability a macroscopic number of particles (clusters of size $O$(1)) or a single cluster completely filling the whole chain. The stationary nonequilibrium phase CF consists of a completely filled chain with current $J = \beta$. The unusual phase transition takes place across the boundary $\alpha = p$ between the MP~I and CF phases.}   \label{PhaseDiag}
\end{figure}

Our extensive Monte Carlo simulations of the model of particles, obeying the above generalized TASEP dynamics, point out to a phase diagram in the ($\alpha, \beta$) plane containing three phases: a many-particle (MP) one, consisting of two subregions MP I and MP II, a completely filled (CF) with particles phase, and a mixed MP+CF phase, see Fig. \ref{PhaseDiag}. First, we discuss the features of the local density profiles which happen to depend essentially on the relative magnitude of the three characteristic probabilities: of  injection $\alpha$, ejection $\beta$, and hopping $p$.

The phase CF ($\alpha \in [p,1]$) represents a chain completely filled with particles, $\rho_i =1$, $ i=1,\dots, L$. This follows from the fact that in the region $\alpha \ge  p$ the modified injection probability $\tilde{\alpha} \equiv  1$ by definition. Hence, at the end of each update, an empty first site is filled deterministically with a particle. The injected particle may hop with probability $p$ to the second site of the chain, whenever that site is empty, thus  leaving the first site ready to be filled deterministically with a new particle in the next discrete-time moment. Thus, a cluster of at least two particles is formed and that cluster grows with time until a stationary state is reached in which all the lattice sites are occupied. Due to the right boundary condition the stationary current of particles is $J = \beta$, which is confirmed by our Monte Carlo simulations.

\begin{figure}
\includegraphics[width=160mm]{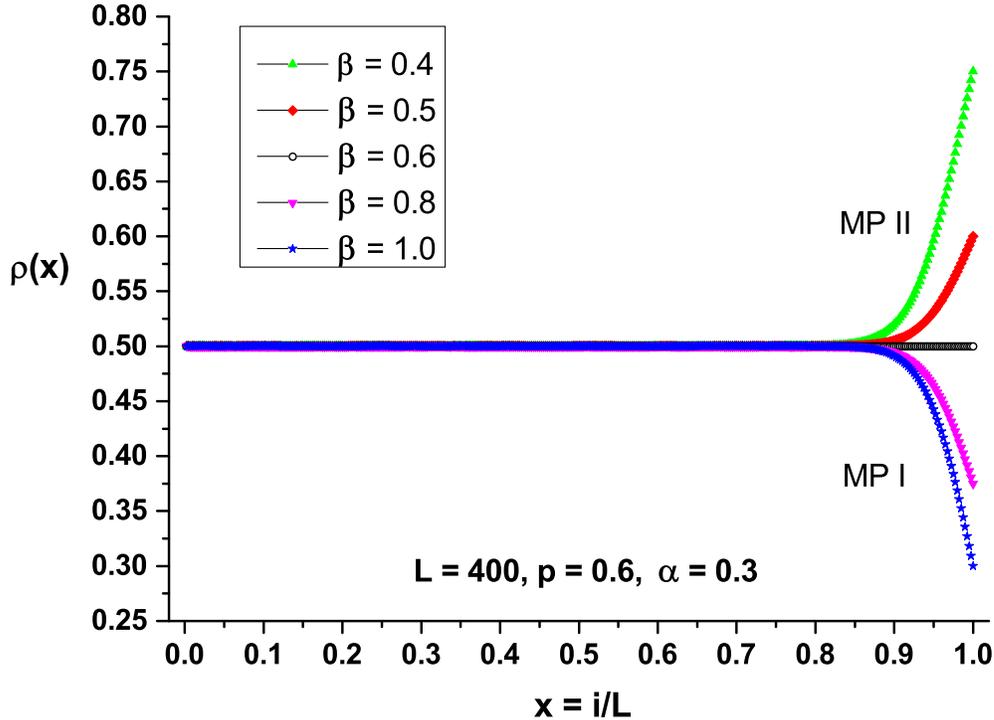} \caption{(Color online) Local density profiles in phase MP for a chain of $L = 400$ sites, hopping probability $p =0.6$, fixed injection probability $\alpha$ = 0.3, and several values of the ejection probability $\beta >\alpha$: $\beta$ = 0.4 (green up triangles), 0.5 (red diamonds), 0.6 (empty circles), 0.8 (magenta down triangles) and 1 (blue stars).}   \label{RoPin}
\end{figure}

The regions MP~I and MP~II ($\alpha < p$  and $\beta >\alpha$) represent a phase containing many particles, or clusters, with bulk particle density $\rho_{\rm b} =\alpha/p$. In contrast to the case of the standard TASEP, where a similar place in the phase diagram is taken by a low-density phase, here the bulk density can take any value from zero to one. The local density profile is flat up to the first site,  $\rho_{\rm 1} =\alpha/p \equiv \tilde{\alpha}$, but the two regions differ by its shape near the chain end where, within numerical accuracy, we find $\rho_L = \alpha/\beta$ see Fig. \ref{RoPin}. Expectedly, on the borderline $\beta = p$ between the regions MP~I and MP~II the local density profile is completely flat:  $\rho_i =\alpha/p \equiv \tilde{\alpha}$, $ i=1,\dots, L$, for all $\alpha \le p$. Analogously, a completely flat density profile occurs in the case of the usual TASEP on the mean-field line: $\alpha + \beta = 1$ for continuous-time kinetics and $(1-\alpha)(1-\beta) = (1-p)$ for discrete-time one.

The phase MP+CF ($\beta < \alpha < p$)  is a mixture of many-particle configurations and nonzero probability of complete filling of the chain in the infinite-size limit. Here, the chain is completely filled at the bulk and up to the last site,  $\rho_{\rm b} =\rho_L =1$. However, the left boundary layer is not completely filled, since the local density $\rho_1$ decreases linearly with $\beta$ down to $\alpha/p$ at $\beta =\alpha$, see Fig. \ref{TransPin}. The particle current is given by $J = \beta$, as in the pure phase CF.

\begin{figure}
\includegraphics[width=160mm]{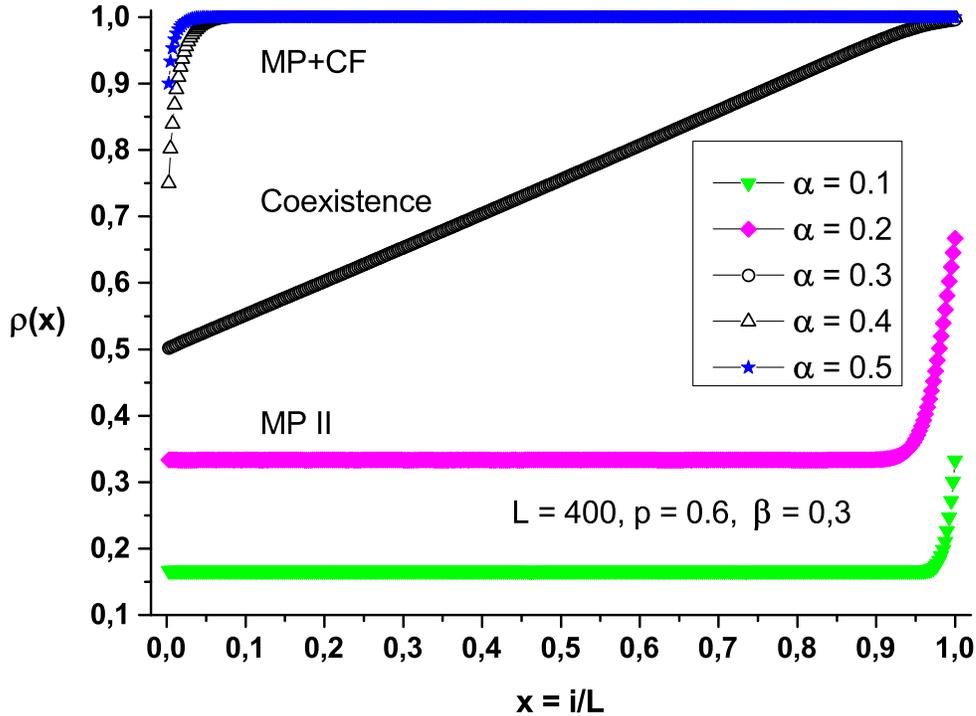} \caption{(Color online) Local density profiles for a chain of $L = 400$ sites, hopping probability $p =0.6$, fixed ejection probability $\beta$ = 0.3, and different values of the injection probability $\alpha$ from phase MP~II: $\alpha$ = 0.1 (green down triangles) and 0.2 (magenta diamonds), through the coexistence line of phases MP~II and MP+CF, $\alpha = \beta$ = 0.3 (empty circles), into phase MP+CF at $\alpha$ = 0.4 (empty up triangles) and 0.5 (blue stars).}   \label{RoOut}
\end{figure}

The typical changes in the local density profiles and the current of particles on passing from phase MP+CF to regions MP~II and MP~I are illustrated in
Fig.~\ref{RoPin}.  On the line $\alpha =\beta$ the density profile is almost linear and can be interpreted as a phase coexistence line between the MP+CF and MP phases, with completely delocalized domain wall, see \cite{KSKS98,SA02}, separating the corresponding particle densities $\rho_{\rm b}^{\rm MP} =\alpha/p$ and $\rho_{\rm b}^{\rm MP+CF}  =1$. The shape of the local density profiles changes with the increase of $\alpha$ from region MP~II to phase MP+CF at fixed $\beta < p$ as shown in Fig. \ref{RoOut}. In contrast, on crossing the borderline between regions MP~II and MP~I nothing changes but the local density profile in the right boundary layer.

\begin{figure}
\includegraphics[width=100mm]{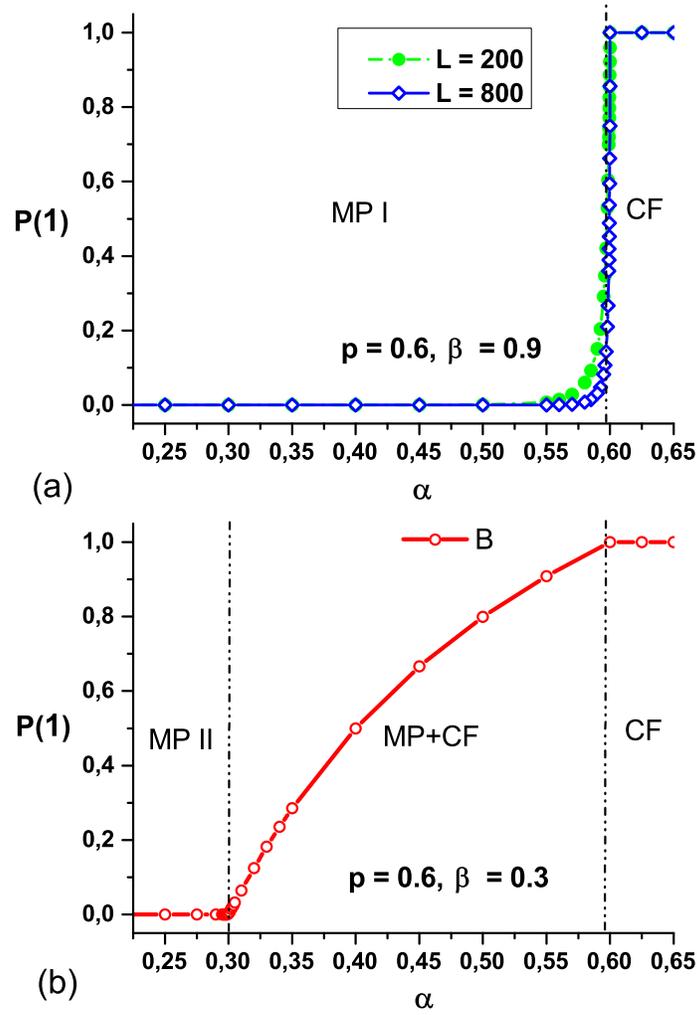} \caption{(Color online) Dependence of the probability P(1) of complete lattice filling in the MF+CF phase close to the boundary with the CF phase. Panel (a) shows that P(1) is identically zero in the bulk of the MP~I phase and grows exponentially fast on approaching the immediate  vicinity of the CF phase. The size-dependence of the distribution is evident from the comparison of profiles for $L = 200$ and $L = 800$, given at ejection probability $\beta = 0.9$. Panel (b) shows the smooth growth of P(1) with $\alpha$ in the MP+CF phase, at $\beta = 0.3$, from P(1) =0 at the boundary with the MP~II phase to P(1) =1 at the boundary with the CF phase.}   \label{C1Dist}
\end{figure}

A distinguishing feature of phase MP+CF, which explanes the name we have given to it, is the behavor of the probability P(1) of finding a completely filled chain with the increase of $\alpha$: it smoothly increases from zero, at the phase boundary $\alpha = \beta$  with the MP~II domain, up to unity at the boundary $\alpha = p$ with phase CF, see Fig. \ref{C1Dist}.

If we interpret  P(1) as an order parameter, we can speak about a continuous,  clustering-type phase transition throughout the domain of MP+CF, from a completely disaggregated phase in MP~II to the completely aggregated one CF.

\section{Phase transitions}

\begin{figure}
\includegraphics[width=160mm]{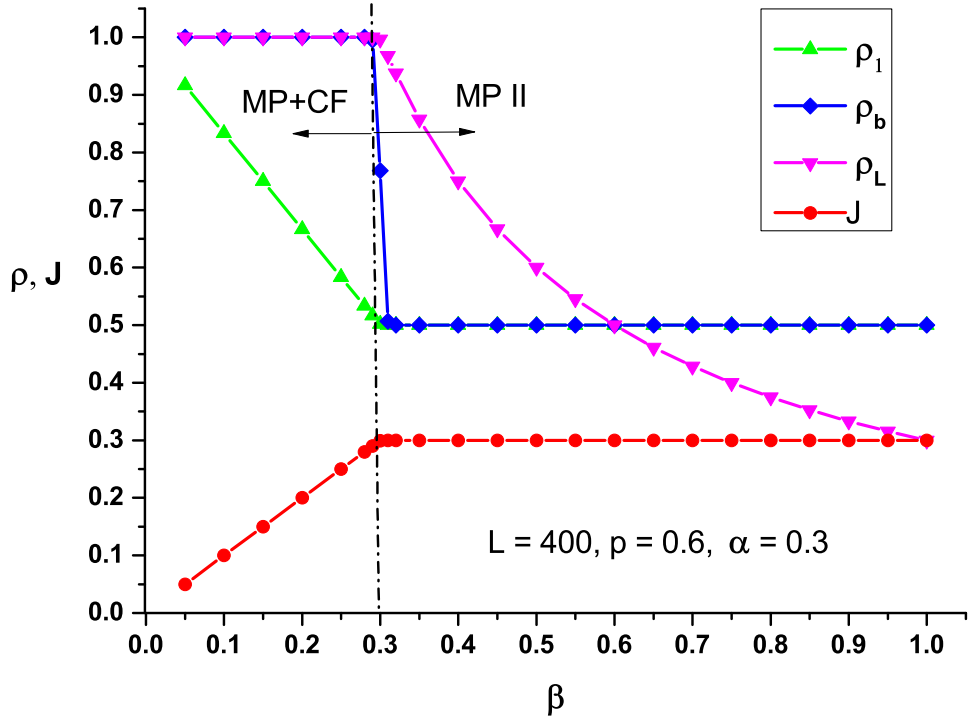} \caption{(Color online) Particle densities $\rho_1$ (green up triangles), $\rho_{\rm b}$ (blue diamonds), $\rho_L$ (red down triangles), and the particle current (red disks) as a function of the ejection probability $\beta$, for a chain of $L = 400$ sites, hopping probability $p =0.6$ and fixed injection probability $\alpha$ = 0.3. The non-equilibrium first-order phase transition at $\beta = \alpha = 0.3$ is clearly manifested by the jump in the bulk density and the discontinuity in the first derivative of the particle current $J(\beta)$. At the borderline $\beta = p$ between regions MP~II and MP~I nothing changes except the tail of the local density profile.}   \label{TransPin}
\end{figure}

From Fig.~\ref{TransPin} it is seen that on passing from phase MP+CF to phase MP with the increase of $\beta$ at fixed $\alpha <p$, a 'jump' in the bulk density from the value $\rho_{\rm b} =1$ down to $\rho_{\rm b} =\alpha/p$ occurs on the borderline $\beta = \alpha$, together with a discontinuity in the first derivative of the current $J(\beta)$. This signals the appearance of a non-equilibrium first-order phase transition in the infinite-chain limit between the phases MP+CF and MP.

\begin{figure}
\includegraphics[width=160mm]{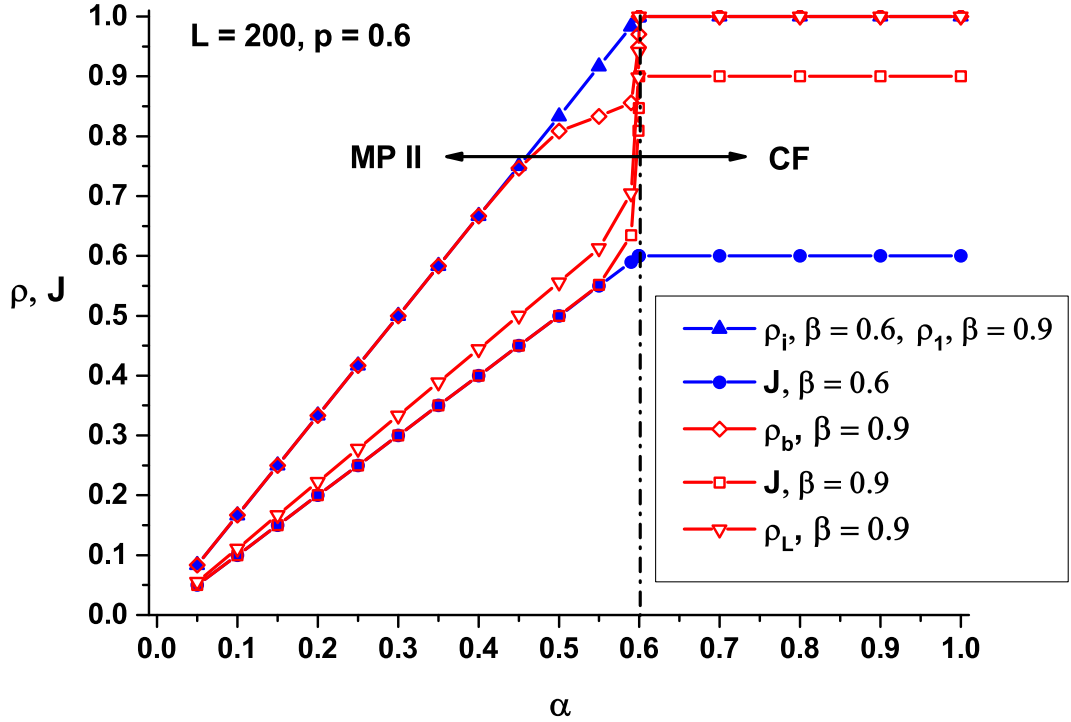} \caption{(Color online) Behavior of the current $J$, bulk density $\rho_{\rm b}$ and local densities $\rho_i$ at site $i$, for a chain of length $L = 200$, in the case of hopping probability $p = 0.6$, as a function of the injection probability $\alpha$ at two values of the ejection probability: $\beta = 0.6$ and $0.9$.  For $\beta = p = 0.6$ the density profile is flat, $\rho_i = \rho_1$, $i = 1,\dots, L$, and equals the local density at the first lattice site $\rho_1$ for $\beta > p$; these values are shown by filled blue up triangles; the current at $\beta = p = 0.6$ is shown by filled blue discs. For $\beta =0.9$ the bulk density $\rho_{\rm b}$, identified with the local density at the center of the chain, $i = L/2$, is shown by empty diamonds with red edges, and the local density at the last site, $\rho_L$ by empty down triangles with red edges; the corresponding current is shown by empty squares with red edges.}   \label{RoJPin}
\end{figure}

On the other hand, a quite unusual non-equilibrium phase transition can be predicted from the $\alpha$-dependence (at fixed $\beta > p$) of the bulk density and the current on passing from phase MP~I to phase CF, see Fig. \ref{RoJPin}. The non-equilibrium 'zeroth-order' phase transition at $\beta >p$  is clearly manifested by the jumps both in the bulk density $\rho_{\rm b}(\alpha)$ and the current $J(\alpha)$ taking place at the boundary $\alpha = p$ between the MP~I and CF phases. Up to our knowledge, a jump in the particle current has never been observed before, at least not in the nonequilibrium stationary states of one-dimensional driven-diffusive systems with conserved bulk dynamics. Informally, we call this phase transition of 'zeroth-order' because in the standard TASEP the second-order transition between the LD (HD) and MC phases is accompanied by discontinuity in the second derivative of the current with respect to $\alpha$ ($\beta$), at the first-order transition between the LD and HD phases there is discontinuity in the first derivative of the current, and here we observe a jump discontinuity in the current itself. Note that across the borderline $\beta = p$ between regions MP~II and MP~I the density and current change continuously, only the curvature of the density profile changes sign.

In region MP I ($0 < \alpha < p < \beta$), within the estimated error bars and for $\alpha$ not too close to $p$, we have estimated
$\rho_1 = \rho_{\rm b} = \alpha/p \equiv \tilde{\alpha}$, $\rho_L  = \alpha/\beta$, $J = \alpha$. Since $\beta >p$, the density profile bends downwards near the chain end, $\rho_L < \rho_{\rm b}$. In region MP II ($0 < \alpha < \beta < p$), again the relations $\rho_1 = \rho_{\rm b} = \alpha/p \equiv \tilde{\alpha}$, $\rho_L  = \alpha/\beta$ and $J = \alpha$ hold true within numerical accuracy. However, since now $\beta >p$, the density profile bends upwards near the chain end, $\rho_L > \rho_{\rm b}$. The above numerical estimates turn out to be exact,  as follows from their theoretical  derivation  in the next section.

A more detailed description of the unusual  'zeroth order'  phase transition is provided by the finite-size analysis of the jumps in the current and bulk density carried out  in Sec. VI.

\section{Derivation of the particle density at the chain ends}

Our numerical results for the local density in phase MP show that the flat profile extends from the first site into the bulk, $\rho_1^{\rm MP} = \rho_b^{\rm MP}$,  and only close to the chain end the profile bends downward in the MP~I domain ($0 <\alpha < p, \beta > p$), upward in the MP~II domain ($0 <\alpha < \beta <p$), and remains completely flat on the line $\alpha = \beta = p$ between these two, see Fig. \ref{RoPin}. Moreover, the numerical results suggest that the particle density in the uniform part of the profile equals
\beq
\rho_1^{\rm MP} =  \alpha/p. \label{MPro1}
\eeq

Here we derive this result from the local balance of the average inflow and outflow of particles at the first site of the chain. Indeed, at the end of a given update, the occupation number of the first site $\tau_1$ changes its value from $\tau_1 =0$ to $\tau_1 =1$ when a particle enters at the empty first site of the system; this event happens with probability $\alpha(1-\rho_1)$. In the opposite case, $\tau_1$ changes from $\tau_1 =1$ to $\tau_1 =0$ when the occupied first site becomes empty; this event happens with probability $(1-\tilde{\alpha})\rho_1 p$. The last expression is valid under the assumption P(1) = 0, when the particle at the first site is either isolated, or belongs to a cluster with rightmost occupied site $j < L$. In this case, under our backward-ordered sequential algorithm the particle will hop with probability $p$ one site to the right alone, if isolated, or with the whole cluster it belongs to, thus leaving the first site vacant. The factor $(1-\tilde{\alpha})$ equals the probability that the site remains empty at the end of the update. Thus, the stationarity condition for the average occupation number of the first site implies $\alpha(1-\rho_1)= (1-\tilde{\alpha})\rho_1 p$, which is equivalent to Eq. (\ref{MPro1}), since $\tilde{\alpha}p = \alpha$. Note that, as shown in Fig. \ref{C1Dist}, the assumption P(1) = 0 generally holds true in the phase MP. Deviations may occur only in a shrinking to zero with the increase of $L$ thin layer at the boundary $\alpha = p$ between the MP~I and CF phases, where P(1) does not vanish.

Now, we pass to the derivation of the average occupation number $\rho_L$ of the last site of the chain in the MP phase. To this end we make use of the global stationarity condition $J_{\rm in} = J_{\rm out}$, where $J_{\rm in}$ ($J_{\rm out}$) is the average inflow (outflow) of particles in the system. First, we prove that a particle enters the system with the same probability $\alpha$, independent of the chain configuration, provided the latter is not completely filled. Consider the two complementary possibilities: the first site is either vacant or occupied by a particle when the backward-ordered configuration update reaches that site. When it is vacant, a particle enters the system with probability $\alpha$ by definition. When it is occupied by a particle, it becomes vacant with the hopping probability $p$ of that particle or the finite cluster containing it. Finally, according to our modified left boundary condition, the first site, that has just become vacant, will be filled by a particle with probability $\tilde{\alpha} = \alpha/p$. Therefore, the total probability with which a particle can enter the system during the update (integer-time moment) is again $\alpha$. Hence,  $J_{\rm in} = \alpha$. On the other side, the outgoing current is by definition $J_{\rm out} = \beta \rho_L$. Hence, the global stationarity condition implies:
\beq
\rho_L = \alpha/\beta.
\label{MProL}
\eeq

Unfortunately, the kinetics in the MP+CF phase is much more complicated due to the presence of a completely filled lattice configuration with probability P(1) depending on both $\alpha$ and $\beta$. This is the reason why we do not present here analogous results for the local particle density at the chain ends in that phase.

\section{Finite-size scaling for the current and bulk density}

\begin{figure}
\includegraphics[width=160mm]{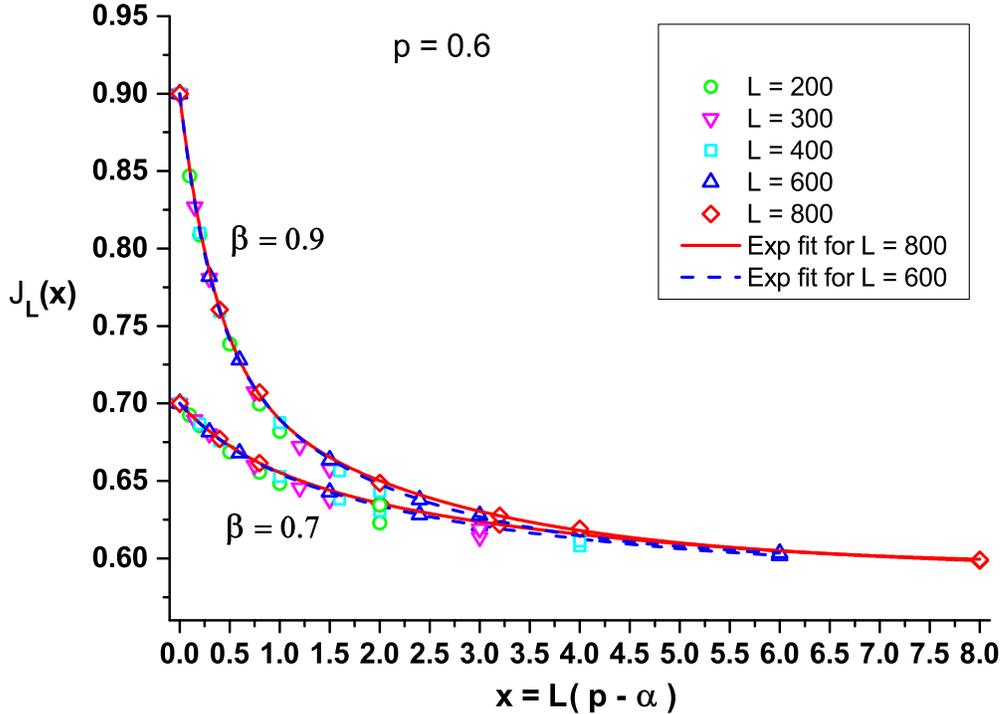} \caption{(Color online) Collapse of Monte Carlo simulation data for the current at $\beta = 0.7$ and $\beta = 0.9$ in chains of different length $L$ as a function of the finite-size scaling variable $x = L(p-\alpha)$: $L = 200$ - green circles, $L = 300$ - red down triangles, $L = 400$ - cyan squares, $L = 600$ blue up triangles, $L = 800$ - magenta diamonds. The two-exponential fits to the data for $L = 600$ is shown by a dashed blue line, and for $L = 800$ - by solid red line, in the cases of $\beta = 0.7$ and $\beta = 0.9$.}   \label{CollapseJ}
\end{figure}

\begin{figure}
\includegraphics[width=160mm]{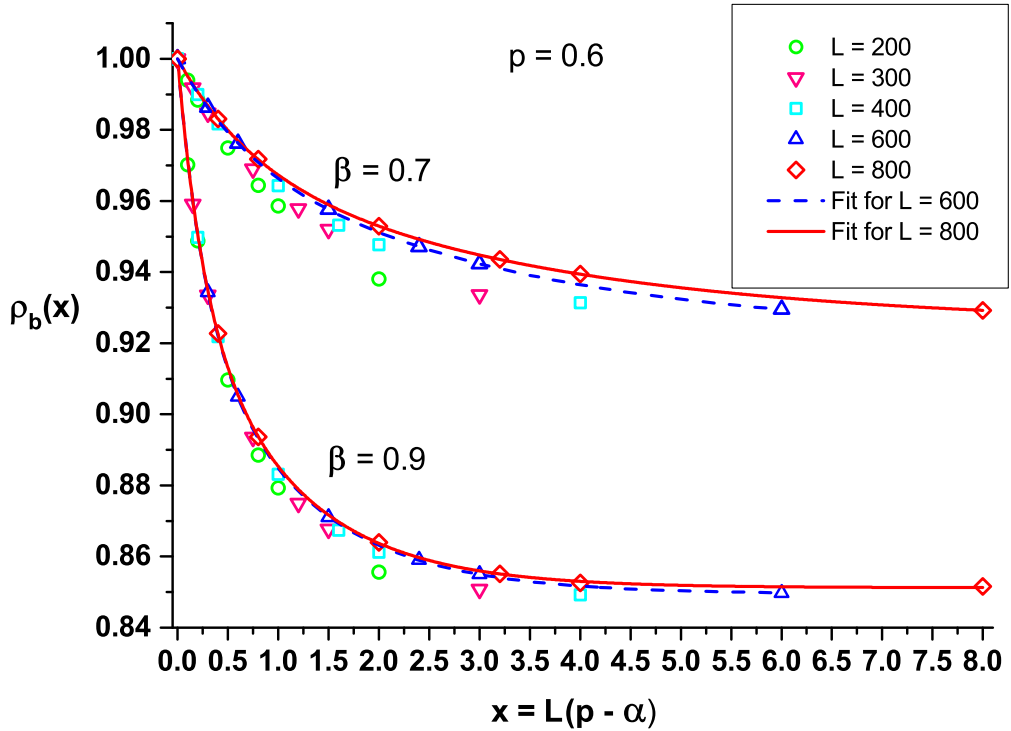} \caption{(Color online) Collapse of Monte Carlo simulation data for the bulk particle density at $\beta = 0.7$ and $\beta = 0.9$ in chains of different length $L$ as a function of the finite-size scaling variable $x = L(p-\alpha)$: $L = 200$ - green circles, $L = 300$ - magenta down triangles, $L = 400$ - cyan squares, $L = 600$ - blue up triangles, $L = 800$ - red diamonds. A two-exponential fit to the data for $L = 600$ is shown by a dashed blue line, and for $L = 800$ - by solid red line, in the cases of $\beta = 0.7$ and $\beta = 0.9$.}   \label{CollapseRo}
\end{figure}

We have attempted a description in terms of finite-size scaling of the sharp changes in both the bulk density and the particle current across the boundary $\alpha =p$ between MP~I and CF from the left. Testing the data collapse method, see, e.g., \cite{NB99,BS01}, under the assumption of a finite-size scaling variable
\beq
x = L(p-\alpha),
\label{FSSv}
\eeq
we have obtained fairly good agreement with the Monte Carlo estimates for both the current, see Fig. \ref{CollapseJ}, and the bulk density, see Fig. \ref{CollapseRo}. Note that we have evaluated the relative accuracy of our simulation data at about $10^{-3}$ for local quantities, and $10^{-4}$ for global ones. These estimates were found by comparing the results obtained under increasing the number of independent runs above $10^6$ per chain site, and changing the initial filling of the lattice. The stationarity of the process was monitored by counting the numbers of injected and ejected particles, which were found to coincide within $10^{-4}$ error.

The data for both the current and the bulk particle density were fitted by the same function
\beq
y(x) = A_1 \exp(-x/\xi_1) +A_2 \exp(-x/\xi_2) + y_0,
\label{fitf}
\eeq
in the cases of $L= 800$ and $L= 600$, at two values of the ejection probabilities $\beta =0.7$ and $\beta = 0.9$, see Figs. \ref{CollapseJ} and \ref{CollapseRo}.

Note that the quality of the fit for the current at $\beta = 0.7$ is excellent: the statistical criteria for $J_L(x)$ are $\chi^2 \simeq 6.5 \times 10^{-11}$, $R^2 = 1$ for $L=800$ and $\chi^2 \simeq 1.9 \times 10^{-8}$, $R^2 = 0.99999$ for $L=600$. At $\beta = 0.9$ the above criteria are somewhat worse, but still satisfactory: $\chi^2 \simeq 3.63 \times 10^{-6}$, $R^2 = 0.99968$ for $L=800$ and $\chi^2 \simeq 1.53 \times 10^{-6}$, $R^2 = 0.99986$ for $L=600$. The values of the corresponding parameters are given in Table I.

The fit for the bulk particle density is also very good: the statistical criteria at $\beta = 0.7$ for $\rho_b(x)$ are $\chi^2 \simeq 7.87 \times 10^{-9}$, $R^2 = 0.99999$ for $L=800$ and $\simeq 1.56\times 10^{-10}$, $R^2 = 1$ for $L=600$. At $\beta = 0.9$ these criteria are again somewhat worse, but sill satisfactory: $\chi^2 \simeq 1.66 \times 10^{-7}$, $R^2 = 0.99994$ for $L=800$ and $\chi^2 \simeq 3.68 \times 10^{-8}$, $R^2 = 0.99999$ for $L=600$. The values of the corresponding parameters are given in Table II. For a robust estimator of the bulk density we have taken the local density at the center of the lattice, $\rho_b = \rho_{L/2}$.

\begin{center}
\begin{tabular}{|c|c|c|c|c|c|c|}
\hline\hline
$\beta$ & L & $A_1$ & $\xi_1$ & $A_2$ & $\xi_2$ & $y_0$
\\ \hline\hline
0.7 & 600 & 0.03687 & 0.7286 & 0.07376 & 3.8393 & 0.58934$\pm 8.5\times 10^{-4}$  \\ \cline{2-7}
    & 800 & 0.03011  & 0.6465 & 0.08052  & 3.2019 & 0.58938$\pm 7.8\times 10^{-5}$  \\ \hline
0.9 & 600 & 0.15613 & 0.31714 & 0.14616 & 1.86285 & 0.59758$\pm 0.0023$ \\ \cline{2-7}
    & 800 & 0.17489  & 0.3581 & 0.1295  & 2.28821 & 0.59546$\pm 0.0031$ \\
\hline\hline
\end{tabular}
\end{center}
\vspace{0.5cm}
Table I. Parameters of the fit (\ref{fitf}) of the collapse data for the particle current, see Fig. \ref{CollapseJ}.

The analysis of the values of the parameters, given in Table I and Table II, allows us to derive important characteristics of the phase transition - the thermodynamic limit of the jumps $\Delta J$ and $\Delta \rho_b$ at the transition from MP~I to CF. Indeed, $y(0) = A_1+A_2 +y_0$ equals the value of the quantity $y$ at the very transition line $\alpha = p$, evaluated for arbitrarily large $L$. On the other hand, $y(\infty) = y_0$ yields the value of $y$ in the thermodynamic limit, arbitrarily close to the transition line on the side of phase MP~I, since $p-\alpha >0$ can be arbitrarily small. Therefore, we obtain
\beq
\Delta y = y(0) - y(\infty),
\label{Delta}
\eeq
where $y$ stays for the current $J$ or the bulk density $\rho_b$.

Now we pass to the evaluation of the thermodynamic jump in the current. From the data in Table I, it can be readily checked that at $\alpha = p$, $\beta = 0.7$, one obtains $J(0)= A_1 + A_2 + y_0 = 0.69997$ at $L = 600$ and $J(0)=0.70001$ at $L = 800$. Obviously, within very high numerical accuracy, this is the $L$-independent value of the current $J = \beta = 0.7$ in phase CF. On the other hand, in the limit $\lim_{L \rightarrow \infty}J(x)$ at fixed $\alpha < p$, $\beta = 0.7$, one obtains $J(\infty) = 0.58934\pm 0,000078$ for $L = 600$ and (almost) the same value $J(\infty) = 0.58938\pm 0.00085$  for $L = 800$. Hence, we estimate the thermodynamic jump in the current at the point $\beta = 0.7$ of the transition line between the MP and CF phases   as $\Delta J = 0.1106 \pm 0.0001$. Similar calculations at the point $\beta = 0.9$ yield $J(0)= 0.89987$ at $L = 600$ and $J(0)=0.89985$ at $L = 800$.
Here the numerical accuracy is lower, but still fairly well agrees with the $L$-independent value of the current $J = \beta = 0.9$ in phase CF. Taking into account the corresponding $y_0$ values from Table I, for the jump in the current at the point $\beta = 0,9$ we obtain the estimate $\Delta J = 0.3035 \pm 0.0035$. Our numerical investigation of the variation of $\Delta J$ along the phase transition line between the phases MP~I and CF (not shown here) has established that $\Delta J$ changes continuously from zero at the triple point $\beta = p$ to unity at $\beta =1$.

In complete analogy with the previous consideration, from the data in Table II, one obtains that at the transition point $\alpha = p$, $\beta = 0.7$, the finite-size scaling function for the bulk density equals $\rho_b(0) = A_1 + A_2 + y_0 = 0.99998$ at $L = 600$ and $\rho_b(0) = 1.00001$ at $L = 800$. Thus, with very high accuracy our data reproduce the value of the particle density $\rho_b = 1$ in phase CF. On the other hand, in the limit $\lim_{L \rightarrow \infty}\rho_b(x)$ at fixed $\alpha < p$, one obtains the estimates $\rho_b(\infty) = 0.92494$, from the data for $L = 600$, and $\rho_b(\infty) = 0.92243$ for $L = 800$. These values, though slightly $L$-dependent, approximate the thermodynamic limit of the bulk density $\rho_b$ in phase MP~I on the left-hand side of the boundary $\alpha = p$ with the CF phase. Hence, the evaluated from the data for $L = 600$ and $L =800$ jump in the bulk density varies from $\Delta \rho_b = 0.07757$ to $\Delta \rho_b =0.07506$ with the final estimate $\Delta \rho_b = 0.0762\pm 0.0015$. Expectedly, the higher is $\beta \in (p, 1]$, the larger is the jump in $\rho_b$. Thus, for the thermodynamic value of the jump at the transition point $\alpha = p$, $\beta = 0.9$ we obtain the estimate $\Delta \rho_b =  0.1496\pm 0.0012$.

\begin{center}
\begin{tabular}{|c|c|c|c|c|c|c|}
\hline\hline
$\beta$ & L & $A_1$ & $\xi_1$ & $A_2$ & $\xi_2$ & $y_0$
\\ \hline\hline
0.7 & 600 & 0.02276 & 0.6571  & 0.05482 & 2.9250  & $ 0.92243\pm 1.02\times 10^{-4}$ \\ \cline{2-7}
    & 800 & 0.02719 & 0.7337  & 0.04785 & 3.3246   & $ 0.92494\pm 4.04\times 10^{-4}$ \\ \hline
0.9 & 600 & 0.06625 & 0.25955 & 0.08426 &  1.09819 & $ 0.84948\pm 2.1\times 10^{-4} $ \\ \cline{2-7}
    & 800 & 0.06157 & 0.24874 & 0.08719 & 1.02839 & $ 0.85124\pm 3.53\times 10^{-4}$ \\
\hline\hline
\end{tabular}
\end{center}
\vspace{0.5cm}
Table II. Parameters of the fit (\ref{fitf}) of the collapse data for the bulk density , see Fig. \ref{CollapseRo}.

It could be instructive to interpret our data-collapse results in terms of the main hypothesis of finite-size scaling at equilibrium phase transitions,
see details in \cite{PF84}. Close to the transition point, the behavior of a singular in the thermodynamic limit quantity in a system of finite size $L$ is described by the variable $L/\xi$, where $\xi$ is a diverging correlation length. For example, it is known \cite{SD93} that the discontinuous with
respect to the bulk density nonequilibrium transition across the coexistence line $\alpha = \beta < \alpha_c = \beta_c$ in the standard TASEP is characterized by a diverging correlation length $\lambda$ of the asymptotic form
\beq
1/\lambda := |1/\lambda_{\alpha} - 1/\lambda_{\beta}| = C|\alpha -\beta| + O\left(|\alpha- \beta|)^2\right).
\eeq
Here $C$ is some model-dependent constant, and
\beq
1/\lambda_{\sigma} = \ln\left[1 + \frac{(\sigma_c -\sigma)^2}{\sigma(p-\sigma)} \right], \qquad \sigma = \alpha, \beta , \quad \sigma_c = 1 - \sqrt{1-p},
\eeq
is the diverging correlation length of the continuous transition from the low-density ($\sigma = \alpha$) or high-density ($\sigma = \beta$) phase to the maximum-current phase, where $1/\lambda_{\rm m.c.} \equiv 0$. Indeed, it was shown that in this case the
finite-size scaling variable is $x_2 = C_2(\beta -\alpha)L$ \cite{B02,BB05}.

From this viewpoint, the choice (\ref{FSSv}) of the finite-size scaling variable for the unusual nonequilibrium phase transition from MP~I to CF at $\alpha = p$ can be interpreted in terms of a diverging correlation length with the asymptotic behavior
\beq
1/\lambda = A(p-\alpha) + O\left((p-\alpha)^2\right),\qquad \alpha \le p,
\label{lambda}
\eeq
where $A$ is some amplitude, possibly parameter dependent. Our data on the cluster-size probability distribution suggests the interpretation of the
diverging correlation length in Eq. (\ref{lambda}) as the size of the largest cluster in the limit of infinite chain $L \rightarrow \infty$. Under this
interpretation the CF phase represents an infinite cluster of particles.

\section{Discussion}

We studied the stationary states of a new one-dimensional model of irreversible aggregation on finite open chains, based on a special discrete-time TASEP-like kinetics. The left boundary condition is set in conformity with the definition of the generalized TASEP on open chains. As a result, the inflow of particles $J_{\rm in} = \alpha$ is independent of the chain configuration, provided it is not completely filled. In addition, the completely aggregated phase occupies a finite domain $\alpha \ge p$ in the plane of particle injection-ejection probabilities ($\alpha$-$\beta$). Since the model has not yet been solved exactly, our study was mainly based on extensive Monte Carlo simulations. By evaluating the local density profiles, the current and the probability of complete lattice filling for different model parameters, we have obtained a phase diagram with a novel topology, see Fig. \ref{PhaseDiag}. Besides the completely aggregated phase CF, two other non-equilibrium stationary phases were distinguished: a many-particle one, MP, and a mixed MP+CF phase with non-vanishing probability P(1) of finding a completely filled configuration. Evidence was found for an unusual discontinuous phase transitions between the MP and CF phases. By using the data collapse method, finite-size scaling was established to hold for the bulk density and the current close to the transition point. The corresponding finite-size scaling variable for the current and bulk density suggested the identification of diverging correlation length for this transition and its interpretation as the size of the largest cluster in the limit of infinite chain. In addition, a conventional first-order phase transition between the phases MP+CF and MP was observed on the line $0<\alpha =\beta < p$. The typical behavior of the local density profile and the current at these transitions was evaluated and illustrated in Figs. 4 and 6. Still another, continuous with respect to the probability P(1), clustering-type transition was observed between the phases MP and CF, throughout the phase MP+CP.

The values of the best fit  parameters in the finite-size scaling function Eq. (4) for the particle current and the bulk density, given in Tables I and II, respectively, allowed us to estimate the value of their jump  in the thermodynamic limit on crossing the boundary $\alpha = p$ between the MP~I and CF phases. Besides the explicitly considered cases of $\beta = 0.7$ and $\beta =0.9$, at $p = 0.6$, we have explored the corresponding jumps in the whole region $p < \beta \le 1$ (not shown here) and found that they both continuously increase as $\beta$ changes from $p$ to unity. Next, the fact that the same function Eq. (4) yields a very good approximation to our finite-size scaling data for both the bulk density and the current, lead us to the hypothesis that the discontinuities of these characteristics at the transition between the MP and CF phases could appear as a result of the same microscopic mechanism. Possibly, this is the finite-size scaling behavior of the probability P(1) of finding complete filling of the chain.

The properties of the stationary non-equilibrium CF and MP+CF phases were shown to be completely different from those of the known TASEP phases. Obviously, this is a consequence of the absence of irreversible clustering properties in the TASEP kinetics, except in the considered limiting case $\tilde{p} =1$ of gTASEP. Next, we succeeded in deriving from simple kinetic stationarity conditions the value of the local particle density at the ends of the chain, as a function of the model parameters in the MP phase. Although the particle density profiles in the MP phase resemble those in the low-density phase of the standard TASEP, we have found in finite chains that a non-vanishing probability P(1) of complete cluster aggregation exists only close to the borderline with the CF phase. The microscopic mechanism of the appearance and behavior of P(1) in the whole MP+CF phase needs a further detailed investigation.

Predictions of different atomistic models of aggregation could be particularly effective in detecting the contributions of specific processes, playing part in real aggregation phenomena. In this way, the study of simple models gives a new insight into the rich world of stationary non-equilibrium phases and the transitions between them. Hopefully, our results may stimulate experimental studies of one-dimensional aggregation of particles in a stationary flow, controlled by the boundary conditions. As a continuation of the present work we intend to take into account cluster fragmentation processes by considering gTASEP with values $\tilde{p} < 1$.

\section*{Acknowledgments}

We are grateful to J. G. Brankov and V. B. Priezzhev for critical reading of the manuscript and fruitful discussions.

\end{document}